# Weighing the techniques for data optimization in a database

ANAGHA RADHAKRISHNAN

Politecnico Di Milano

Milan, Italy

Anagha.R@mail.polimi.it

**Abstract**

A set of preferred records can be obtained from a large database in a multi-criteria setting using various computational methods which either depend on the concept of dominance or on the concept of utility or scoring function based on the attributes of the database record. A skyline approach relies on the dominance relationship between different data points to discover interesting data from a huge database. On the other hand, ranking queries make use of specific scoring functions to rank tuples in a database. An experimental evaluation of datasets can provides us with information on the effectiveness of each of these methods.

**Keywords :** Skyline, Top-k query, regret minimizing set, ranking queries

## 1 INTRODUCTION

Determining the most interesting data from a large dataset is an important task and has many important applications in real world scenarios. This becomes especially useful in multi-criteria decision making where the best data has to be chosen from a set of conflicting criteria represented by attributes of tuples in a data-set according to user preferences.

Here we discuss three mostly used approaches: 1)skyline approach: which returns all the non- dominated tuples from a set of records in a database. 2)skyline ranking approach: where a multi-objective problem is reduced to single-objective problem using a scoring function. 3) Regret Minimization: This approach filters out a set of most important tuples from a large database based on users criteria using a set of utility or scoring functions and tries to minimize the maximum regret ratio. Users can input utility functions.

The idea of flexible skyline queries is discussed to understand how user preferences are expressed by using constraints on the weights of a limited set of specific scoring functions.

An example scenario would be a customer booking a vacation stay for a cheap price and close to the airport. Here the main attributes are price and distance. There should be a trade-off between these attributes to make a decision based on the customer's satisfaction. The skyline queries may develop a large set of interesting results and may not express user preferences as attributes which have the same importance. F-skylines can take into account the different importance of different attributes and can model user preferences by means of constraints on the weights used in a scoring function. This results in the reduction of the result size compared to the traditional skyline. A k-regret query outputs k tuples from the database and tries to minimize the maximum regret ratio.

Table 1 represents a set of eleven vacation stays with their price for a day and closeness to the airport. Each row in the table gives information to identify the stay preferred by the customer. The two numeric attributes considered for the customer are price and distance. We consider the criterion to minimize the price and minimize the distance from the airport.

|  | Price(ineuro) | Distance(in km) |
|---|---|---|
| S1 | 250 | 1400 |
| S2 | 200 | 1200 |
| S3 | 400 | 1300 |
| S4 | 550 | 200 |
| S5 | 500 | 100 |
| S6 | 150 | 1300 |
| S7 | 250 | 1000 |
| S8 | 650 | 200 |
| S9 | 350 | 300 |
| S10 | 400 | 600 |
| S11 | 550 | 1000 |

Table 1: Dataset of stays

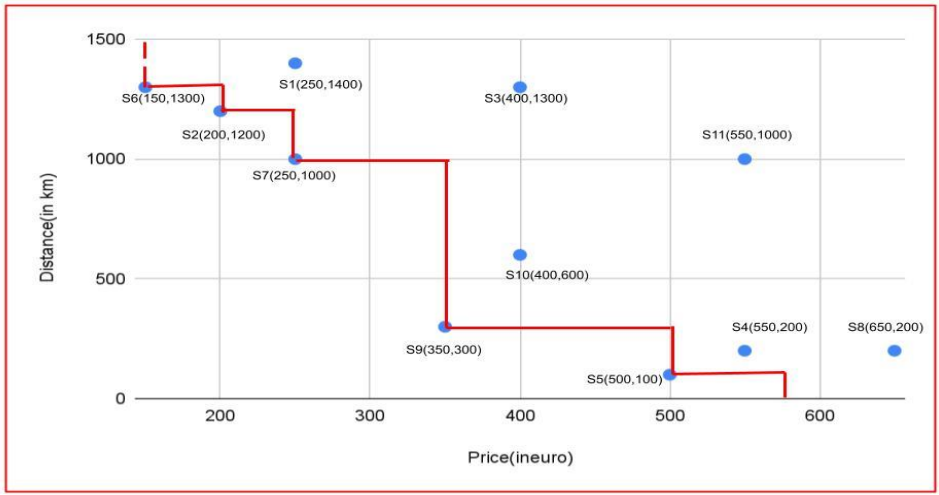

Figure 1: Skyline of set of vacation stay

This paper is a comparative survey of these techniques for getting the important tuples from a database of large dimensions. We mainly focus on the basic idea of the approaches and discuss its limitations and compare its advantages over the other.

## 2  SKYLINE QUERIES

Skyline queries are able to find interesting non-dominated data points from multidimensional datasets. Given a dominance relationship in a dataset, a skyline query returns all the non-dominated tuples resulting in an increasing set of outputs. They cannot accommodate user preferences or cannot control the cardinality of the result set. All the attributes have the same priority as user preferences are not considered and this makes it difficult to choose from.

An example skyline query is as follows:

SELECT * FROM Users
SKYLINE OF age MIN

 A 'SKYLINE OF' clause is added to select statements of SQL query. Dimensions and aggregation functions can be also used for skyline query.

The arbitrary size constraints on skyline queries are resolved using skyline ordering approach. A skyline ordering approach introduces a skyline-based partitioning of a given dataset using the dominance relationship in skyline.

## 2.1 Limitations of skylines queries

As discussed, traditional skyline queries cannot take into account user preferences and cannot control the cardinality of the result set. Operation costs of skyline queries are high. All the attributes have the same priority as user preferences are not considered and this makes it difficult to choose from. The enormously increasing size of the result set also makes skylines less attractive. Therefore, a more effective skyline approach called Flexible skylines (F-skylines) is introduced.

## 3 FLEXIBLE SKYLINES

Flexible skylines (F-skylines) is an effective approach for multi-criteria optimization using different attributes for decision making processes concerned with the data.

Flexible skyline queries extend the concept of traditional skyline framework by incorporating user preferences expressed by using constraints on the weights of a limited set of specific scoring functions. The weights of the scoring function are learned via crowd-sourcing tasks which involve active and direct user involvement. This is achieved by using the dominance concept on scoring function. (F-dominance means dominance with respect to a family of scoring functions F).

Based on the F-dominance concept, two flexible skyline operators were introduced: ND; a set of non-dominated tuples and PO; a set of tuples that are potentially optimal. ND is a subset of the skyline and PO is a subset of ND.
As the constraints become restrictive, the results of ND & PO also get reduced. While ND and PO coincide and reduce to the traditional skyline when F is the family of all monotone scoring functions.

### 3.1 Computing ND and PO

ND is a subset of the skyline. So, to compute ND, we can actually compute the skyline and obtain the result from it. Alternatively, ND can also be computed directly from the input dataset. For the first approach, any skyline algorithms like BNL (Block Nested Loop) and SFS (Sort Filter Skyline) can be performed. SFS performs a preliminary topological sort of the input data. The need to sort/ not sort input data can be changed if a single-phase approach is used (ie, without first computing the skyline).

To compute PO, we can compute ND as an intermediate result or directly obtain the potentially optimal tuples from the input data. Another approximation strategy in which we discard non-optimal tuples can also be used.
Another method involves the use of the primal or the dual PO test.

Efficiency of the results is based on the efficiency of the algorithm. Efficiency of algorithms can be affected by the factors like execution time, number of dominance tests, number of F-dominance tests. Performance of our algorithms clearly depends on the selectivity of the constraints.
The effectiveness of F-skylines steadily improves as the number of constraints increases. Each constraint reduces the space of weights, thus reducing the set F of functions to consider for F -dominance and the number of retained tuples.

## 3.2 Advantages of using F-skylines

- Usage of user preferences on a set of attributes.
- Computation of the F-dominance region of a tuple can be performed just once, thus independently of how many F dominance tests are involved. This reduces the cost of calculation of F-dominance regions rather than calculating for each tuple.
- F-skylines are tailored to deal with numeric attributes.
- Thus, F-skyline operators have good flexibility in controlling the size of the result.
- Computing Skyline and ND via a one-phase algorithm reduces the execution time to a great extent.
- Computing PO incurs only a moderate overhead for all the considered datasets with default parameter values.

## 3.3   Limitations of skyline operator

- In case of dominance, reporting same result for every user (no personalization)
- For quantification or attributes, a preference input is required for practical decision support.
- But for design considerations (like display size, device capabilities, connection speed), a proper output size is required. So, there is a need for output-size specified (OSS) operators.

## 3.4 Overcoming the limitations

To achieve personalization, linear scoring principle is used for decision making.
For decision support, we consider 3 requirements: Personalization, controllable output size (OSS), flexibility in preference specification. From this we propose 2 new operators: ORD & ORU.

ORD/ORU techniques require no precomputation. The ORD algorithm relies on a progressive
The ORU algorithm is based on ranking by utility.

The ORD reports the k skyline members that dominate the most non-skyline records. ORD has the ability to scale. ORU also scales well with all parameters. The running time of ORU increases faster than ORD because, despite the similarities in their definition, the nature of ORU is significantly more complex. Higher dimensionality affects ORU's performance more than ORD's.

## 3.5 P-skylines (prioritized skylines)

User preferences are integrated by introducing a partial order of priorities between attributes. P-skyline approach essentially combines lexicographic and skyline queries and allows no trade-off between attributes.

## 4   REGRET MINIMIZATION

K-representative regret minimizing query (k-regret) is a useful operator for supporting multi-criteria decision-making. This approach filters out a set of most important tuples from a large database based on users criteria using a set of utility or scoring functions. For any 'k' class of utility functions, k-regret query outputs k tuples from the database and tries to minimize the maximum regret ratio. The regret ratio decreases monotonically with k. Users can input utility functions. We consider the maximum regret ratio as a NP hard problem.

### 4.1 K-regret operator

K-regret operator has features of both top-k and skyline. This operator outputs a small set of k tuples without user utility functions and minimizes the maximum regret ratio. We consider the maximum regret ratio as a NP hard problem.

### 4.2 Properties of K-regret operator

- Scale invariant: Even if we multiply values in some attributes with a constant, k-regret still outputs the same number of tuples. Rescaling attributes does not change relative values of the solutions.
- Stable: Adding an unimportant info into db does not change the solution of k-regret. A function is stable if it is insensitive to adding/deleting junk points.

### 4.3 Advantages of K-regret approach

- Maximum regret ratio is bounded and this bound is independent of database size.
- Maximum regret ratio is reasonably small.
- The algorithms related to this approach run in linear time in the size of the database and the running time is small.
- Minimizing maximum regret ratio takes negligible time as compared to running skyline computation.
- Due to the small output size, the algorithm runs faster than a huge skyline.

### 4.4 Limitations of K-regret approach

Inputting weights to each criterion. We can try to reduce the number of criteria based on users input. But the size of the skyline could be very large even when only some criteria are involved. As the number of criteria increases, skyline size grows exponentially.

### 4.5 K-regret minimizing sets (kRMS)

Representing an entire dataset by a few representative points for multi-criteria decision making. An optimal solution if found using algorithms like dynamic programming or greedy randomized algorithms based on LP. We consider the k-regret minimizing set that represents the dataset not by approximating the user's top-1 choice, but by taking into account their top-k choices thus making the solution quality much stronger.

We produce a fixed-size k-regret minimizing set that achieves the minimum possible maximum k-regret ratio. We aim for a fast algorithm to find sets with a low k-regret ratio. Real dataset with normalized constraints is considered based on solution quality and execution time. The impact of randomization on the solution quality is considered and missing values have been replaced with the lowest value found in the dataset. Non-skyline points have been pruned, because they will never form part of any solution.

### 4.6 Advantages of kRMs
- Introduction of regret minimizing sets addressing the computational issues.
- The dual-space top-k rank contours which answer monochromatics reverse top-k queries.
- kRMS hybridizes the skyline with top-k queries.
- Produces output of the exact desired size.

### 4.7 Interactive regret minimization (IRM)

IRM is obtained by including personalization in RMS. It involves the user in the search process through multiple rounds of interaction. The record which points to the minimum regret ratio is then selected for the user. So there is active user involvement. Its objective is to eventually identify the one record with maximum utility, and thus considers only records on the skyline.

## 5 SKYLINE RANKING

In this approach, the original multi-objective problem is reduced to a single objective problem using a scoring function. Function parameters like weights are used to adjust scales and priority is given to attributes by the user using the scoring function. Results of ranking queries heavily depend on how user preferences are translated into a particular choice of weights in the scoring function. They adopt a specific scoring function to rank tuples and can easily control the output size.

### 5.1 Skyrank

A framework for ranking the skyline points in the absence of a user-defined preference function. A limited set of interesting points of the skyline set are discovered.

### 5.2 Skyline graph

A graph which relies on the dominance relationships between the skyline points for different subsets of dimensions (subspaces).

Skyrank applies well-known authority-based ranking algorithms on the skyline graph. This also discovers the importance of a skyline point exploiting the subspace dominance relationships. Skyrank can be extended to handle top-k preference skyline queries, when the user's preferences are available. Using this method, we are able to apply link-based techniques to rank the skyline points. Advantages of kRMs

### 5.3 PageRank

Page Ranking is a link-based technique for ranking and evaluates the ability of the dominance relationship to provide a meaningful ranking. It assigns a score to each vertex, i.e. skyline point. This scores are then used for ranking and answering top-k skyline queries. The PageRank score for page p is defined as the stationary probability of finding the random surfer at page p and relates to the number and the importance of the pages pointing to it.

## 5.4 Skyline frequency

It compares and ranks the interesting of data points based on how often they are returned in the skyline, when different subspaces are considered. It is a metric for ranking the skyline points. A point with high skyline frequency is more interesting as it can be dominated on fewer subspaces.

## 5.5 Limitations skyline frequency

All skyline points which appear in the same subspaces have the same skyline frequency.

## 5.6 Telescope algorithm

Ranks the skyline points by user-specified preferences on the available dimensions.

## 5.7 Advantages of skyline ranking

- Easily adapted to user-defined preferences
- SKYRANK supports top-k preference skyline queries by using the personalization vector of the PageRank computation. .

## 6 CONCLUSION

In this paper, we have compared the most useful and effective techniques for multi-criteria decision making from a large set of data. Each of the techniques have been introduced and a general idea on the methods are given. We discussed the importance of the techniques and also discussed their advantages and limitations. The limitations traditional skylines which are selection of user preferences and controlling cardinality of the result set gives motivation to the use of flexible skylines. Adding to these advantages k-regret approach filters out a set of k most important tuples from a large database based on users criteria using a set of utility or scoring functions and tries to minimize the maximum regret ratio.

Skyline ranking approach on the other hand make use of function parameters like weights used to adjust scales and priority is given to attributes by the user using the scoring function. Thus they depend on user preferences and also controls the output size.